\documentclass[twocolumn,nofootinbib]{revtex4}%
\usepackage{amssymb}
\usepackage{amsfonts}
\usepackage{amsmath}
\usepackage{graphicx}
\usepackage{color}

\begin{document}

\title{Rotating hairy black hole and its microscopic entropy in three spacetime dimensions}
\author{Francisco Correa,${}^1$ An\'{i}bal Fa\'undez,${}^{1,2}$ and Cristi\'an Mart\'{\i}nez${}^{1,3}$}
\email{correa, faundez, martinez@cecs.cl}
\affiliation{
 ${}^1$Centro de Estudios Cient\'{\i}ficos (CECs),  Av. Arturo Prat 514, Valdivia, Chile\\ 
${}^2$Departamento de F\'{\i}sica, Universidad de Concepci\'on, Casilla 160-C, Concepci\'on, Chile \\
${}^3$Universidad Andr\'es Bello, Av.\ Rep\'{u}blica 440, Santiago, Chile
}

\begin{abstract}

We present a spinning hairy black hole in gravity minimally coupled to a self-interacting real scalar field in three spacetime dimensions. The black hole is characterized by having a single horizon which encloses a curvature singularity and the scalar field is regular everywhere. The mass and angular momentum are shown to be finite. The presence of a scalar field with a slower fall-off at infinity leads an anti-de Sitter asymptotic behavior which differs from the one found by Brown and Henneaux, but has the same symmetry group as in pure gravity. A scalar soliton, which is a finite mass regular solution devoid of integration constants, plays the role of the ground state. The existence of this soliton is the key to derive the semiclassical entropy of the rotating hairy black hole using the counting of microscopic states provided by the Cardy formula.

\end{abstract}
\maketitle

\section{Introduction}

Three-dimensional gravity has been widely considered as a useful laboratory for the understanding of gravitation in four and higher dimensions. One of the most important examples is given by the Ba\~nados, Teitelboim and Zanelli (BTZ) black hole \cite{BTZ}, which illustrates several features of higher dimensional black holes and it  has been shown to be  relevant in different contexts,  like the AdS/CFT correspondence, among many others. Concerning the holographic principle approach,  a strong evidence was posed earlier by Brown and Henneaux  in the case of gravity with negative cosmological constant in three dimensions \cite{Brown-Henneaux}. They shown that the asymptotic conditions of three-dimensional anti-de Sitter spaces (AdS) are left invariant under the conformal group in two dimensions. This means, using the AdS/CFT correspondence, that this conformal field theory,  defined on  a cylinder at spatial infinity, should correspond to the quantum AdS gravity in three dimensions.  A direct application of the correspondence is the microscopical computation of the semiclassical black hole entropy by means of Cardy formula, as was done by Strominger \cite{Strominger} for the  BTZ black hole. Relevant discussions and further approaches in this subject have been provided by Carlip in \cite{Carlip}. Recently, several attempts had been performed in order to compute the black hole entropy using a field theory at the boundary in different three-dimensional theories. For instance, gravity with scalar fields \cite{CMT,CMT2}, new massive gravity \cite{BHT1} with AdS \cite{Giribet:2009qz,PTT} and Lifshitz \cite{GTT} asymptotics, higher spin theories \cite{PTT2},  gravity with asymptotically flat cosmological solutions \cite{4371,4372} and topologically massive gravity \cite{DHH}. 
 
The entropy of static hairy scalar black holes has been computed, by means of  Cardy formula, in the case of potentials defined by one \cite{HMTZ-2+1,CMT} and two coupling constants \cite{CMT2}. The essential point to achieve that relies on the suitable choice of the corresponding ground state, which turns out to be a scalar soliton instead of AdS spacetime. In fact, AdS spacetime works properly as a ground state \textit{only} for the pure gravity sector. Therefore, it was necessary to find the proper gravitational dual of the ground state, being the mass of this configuration the lowest eigenvalue of the zero-mode Virasoro operator. 

In this article we provide a new class of spinning hairy black holes and show that a scalar soliton is again essential in order to reproduce its semiclassical entropy by using the asymptotic growth of the number of states. Rotating scalar hairy black holes coupled to different classes of gravity theories in three dimensions have been found in the past  \cite{ChanMann, LemosSa, Natsuume}. Here,  we focus our attention in a particular class of spinning hairy black holes, which can be built from a static hairy black hole by means a Lorentz boost in the plane $t-\varphi$. The black hole solution is an asymptotically  AdS spacetime, in a relaxed form in comparison with Brown-Henneaux (BH) conditions, and has a single horizon, which encloses a singularity located at the origin. The scalar field is regular everywhere, endowing the black hole with a scalar hair. 

In the next section the rotating hairy black hole and the analysis of its geometrical features are presented. Section \ref{sec3} is devoted to compute the mass and angular momentum using the Regge-Teiltelboim approach, and also contains the discussion of the thermodynamical properties. In section \ref{sec4} we obtain the hairy black hole entropy by means of Cardy formula. Finally, some concluding remarks are given in the last section.

\section{Rotating hairy scalar black hole in three dimensions}\label{sec2}
We focus our attention in Einstein gravity minimally coupled to a single real scalar field, with a self-interaction potential,  in three spacetime dimensions. The system is described  by the action
\begin{equation}
\! I[g_{\mu\nu},\phi]=\frac{1}{\pi G}\int d^{3}x\sqrt{-g}\left[  \frac{R}
{16}-\frac{1}{2}(\nabla\phi)^{2}-V(\phi)\right], \label{Action}
\end{equation}
where the self-interaction potential is given by
\begin{equation} \label{potential}
V(\phi)= -\frac{1}{8l^{2}} \cosh^{6}\phi. 
\end{equation}
The physical properties of this potential and the role of the parameter $l$ can seen as follows. For a small scalar field, $\phi <<1$, the potential behaves as
\begin{equation} \label{potential2}
V(\phi)=-\frac{1}{8 l^{2}}-\frac{3}{8 l^{2}}\phi^2 -\frac{1}{2 l^{2}}\phi^{4} +O(\phi^6),
\end{equation}
indicating the presence of a negative cosmological constant  $-1/l^2$ and also of a negative mass term $-3/(4l^2)$ which, however, satisfies the Breitenlohner-Freedman bound in three dimensions. It has been shown \cite{HMTZ-2+1, CMT, CMT2} that a potential behaving as (\ref{potential2}) induces a slower fall-off on the scalar field,  and consequently on the metric,  in the asymptotic region.  Hence, nonlinear terms contribute to the mass and angular momentum.

The field equations admit the following solution
\begin{equation}\label{metricnucero}
 ds^{2}=-N^2(r)dt^2+
\frac{dr^2}{F(r)}+R^2(r)\left(d\varphi +N^{\varphi}(r)dt \right)^2,
\end{equation}
where
\begin{eqnarray}\label{fgh}
N^2(r)&=&\frac{(r-B)(r+B)^4(1-\omega^2)}{l^2((r+B)^3-\omega^2(r+2B)^2(r-B) )}, \\  F(r)&=&\frac{(r+2B)^4(r-B)}{l^2(r+B)^3}, \\
R^2(r)&=&\frac{(r+B)^4-\omega^2(r+2B)^2(r^2-B^2) }{(1-\omega^2)(r+2B)^2},\\ 
N^{\varphi}(r)&=&\frac{\omega B^2 (3r+5B) }{l( (r+B)^3-\omega^2(r+2B)^2(r-B) )}, \label{ocho}
\end{eqnarray}
and the scalar field is given by
\begin{equation}\label{sca}
\phi(r)={\rm arctanh}\, \sqrt{\frac{B}{r+2B}}.
\end{equation}
The solution contains two integration constants, $B$ and $\omega$, and the coordinates range as  $-\infty <t< \infty $, $-B <r< \infty $,  and $0 \leq \varphi< 2\pi$. The constant $B$ must be positive in order to have a real scalar field in the asymptotic region, i.e.  for a large $r$. The meaning of $\omega$ will be clarified below.

The Ricci scalar
\begin{equation}
R^{\mu}_{\mu}=-\frac {2 (r+2 B )^3\left (4 B^2 + 5 B r +  3 r^2 \right)} {l^2 (r+B)^5}
\end{equation}
reveals that there is a single curvature singularity located at $r=-B$ (at $r\to \infty$ approaches to $-6/l^2$) and therefore the radial coordinate starts from there, $r>-B$. The same curvature singularity appears in the invariants $R_{\mu \nu }R^{\mu \nu }$ and $R_{\mu \nu \lambda \rho  }R^{\mu \nu  \lambda \rho}$.

As is very well known, it is a big  challenge to solve the field equations associated to the action (\ref{Action}), even in the static case which could be considered the simplest one. However, we have been informed that a general method can be developed \cite{Grumiller}. We adopt here a procedure that has been proven to be useful for other systems in the past. In three spacetime dimensions,  it is possible to obtain a stationary and axisymmetric solution by applying a \textit{improper} gauge transformation to a static solution in the plane $(t,\varphi)$ \cite{Stachel}. Using this procedure one can get the rotating BTZ black hole, as well as,  the electrically charged version of that \cite{charged}, from the corresponding static solutions. Other examples include black holes with a dilaton field \cite{LemosSa} and with a conformally coupled scalar field \cite{Natsuume} (see also \cite{clement}) .   

Thus, if  one considers the static hairy black hole of Ref. \cite{HMTZ-2+1}
\begin{equation}\notag 
ds^{2}\!=\!-\frac{(r^2-B^2)}{l^2 }d\tilde{t}^2+
\frac{l^2(r+B)^3 dr^2}{(r+2B)^4(r-B)}+\frac{(r+B)^4 }{(r+2B)^2}d\tilde{\varphi}^2,
\end{equation}
which has the same scalar field given in (\ref{sca}), and  performs the following boost 
\begin{equation}\label{boot}
\tilde{t}=\frac{1}{\sqrt{1-\omega^2}}(t+\omega l \varphi), \quad 
\tilde{\varphi}=\frac{1}{\sqrt{1-\omega^2}}(\varphi +\omega t/l ),
\end{equation} 
where the constant $\omega$ is restricted to satisfy $\omega^2<1$, one directly obtains the rotating hairy solution described  by Eqs. (\ref{metricnucero})-(\ref{sca}).  This transformation becomes the identity by setting $\omega=0$.

Taking the advantage that the line element (\ref{metricnucero}) is written in the ADM form, it is easy to go to
 Eddington-Finkelstein coordinates and show that this solution has an event horizon located at $r_+=B$. For this purpose it is necessary to study the existence of poles in the functions $N^2(r)$ and $F(r)$. The lapse $N^2(r)$ is a regular function, free of poles, provided $\omega^2<1$. In this case, the lapse is a positive and monotonically increasing function for $r>r_+=B$.  The function $F(r)$ has no poles for $r>-B$. Therefore, $\omega^2<1$ appears as a necessary condition, not only for keeping a well-defined boost, but to get the spinning black hole from the static one.  
Hence, this stationary and axisymmetric solution corresponds to a hairy rotating black hole, where an event horizon defined at $r=B$ encloses the singularity at the origin $r=-B$.  The scalar field is regular everywhere and remarkably, in contrast with the rotating BTZ black hole, the solution has no an inner horizon like. Indeed, the curvature singularity prevents the existence of a inner horizon for this rotating black hole.

The asymptotic behavior of the solution  (\ref{metricnucero})-(\ref{ocho}) and (\ref{sca}) is given by  	
\begin{equation}
\phi=\left(\frac{B}{r}\right)^{\frac{1}{2}}-\frac{2}{3}\left(\frac{B}{r}\right)^{\frac{3}{2}}+O(r^{-5/2})
\label{asympt scalar general}%
\end{equation}%
\vskip -10pt 
\begin{alignat}{2}\notag
g_{tt}&=-\frac {r^2} {l^2}   + O(1), \,\,\,   g_{rr}=\displaystyle \frac {l^2} {r^2} - \frac {4 B l^2 } {r ^3} + \frac{15 B^2 l^2} {r^4} + O(r^{-5})&\\
g_{\varphi\varphi}&=r^2 + \frac {2 + \omega^2} {1 - \omega^2}B^2 +O(r^{-1}) \notag \\ 
\displaystyle g_{t\varphi}&=\frac {3 \omega B^2 } {l\left (1 - \omega^2 \right)}+
O(r^{-1}) \label{asympt metric general}
\end{alignat}
and satisfies the asymptotic conditions of Ref. \cite{HMTZ-2+1}. Note that this asymptotic behavior is relaxed in comparison with the BH conditions.

\section{Mass, angular momentum and thermodynamics}\label{sec3}

The relaxed asymptotic conditions of  Ref. \cite{HMTZ-2+1}  turns out to be left invariant under the conformal group in two dimensions which is spanned by two copies of the Virasoro algebra. As a consequence, the scalar field contributes to the generators of the asymptotic symmetries, which can be found using the Regge-Teitelboim approach \cite{Regge-Teitelboim}. In fact, by using Eq. (22) of Ref. \cite{HMTZ-2+1} (see also \cite{CMT} and \cite{Jack})
the mass $M$ and the angular momentum $J$ of the hairy black hole described by Eqs. (\ref{metricnucero})-(\ref{sca}) are found to be 
\begin{equation}\label{MJ}
M=\frac{3B^2(1+\omega^2)}{8Gl^2(1-\omega^2)} \, , \qquad J=\frac{3B^2 \omega}{4Gl(1-\omega^2)} \, ,
\end{equation}
which depend on the integration constants $B$ and $\omega$. The reference background, defined  by setting $B=\omega=0$, corresponds to the massless BTZ black hole. As is expected, for $\omega=0$ the angular momentum vanishes and the mass coincides with the one corresponding to the static case. The mass is non-negative ($\omega^2<1$) and the angular momentum has the same sign as $\omega$. Moreover, $M > |J/l|$.

Is straightforward to write the integration constants $B$ and $\omega$ in terms of the mass and angular momentum. In this way, the rotating hairy black hole solution is written just in terms of $M$ and $J$. In particular, it is interesting to observe that the horizon area $A=2\pi R_+$, where $R_+$ is given by
\begin{equation}
R_+^2=\frac{64G l^2}{27}\left(M+\sqrt{M^2-\frac{J^2}{l^2}} \right) ,
\end{equation}
has the same form (up to a numerical factor) as in the case of the rotating BTZ black hole. One can observe that in the limit $M=|J/l|$ the scalar field vanishes and the metric becomes that of the extreme BTZ black hole (see Ref. \cite{charged} for further discussion about this limit).

The temperature of the rotating hairy black hole can be obtained by means of the surface gravity $\kappa^2=-\frac{1}{2}(\nabla^a \chi_b )(\nabla_a \chi^b )$, where the Killing vector $\chi^a $
equals $(1,0,-\Omega)$ and $\Omega \equiv d\varphi/dt=\omega/l $ is the angular velocity at the horizon. Then, the temperature $\kappa/(2\pi)$ reads
\begin{equation} \label{temp}
T=\frac{3 \sqrt{3 G M}}{2\pi l}\sqrt{\left(\frac{M^2 l^2}{J^2}-1\right)\left(1-\sqrt{1-\frac{J^2}{M^2 l^2}} \right)}.
\end{equation}
The Bekenstein-Hawking entropy is defined in terms of the horizon area
\begin{equation}\label{BHentro}
S=\frac{A}{4G}=\pi \sqrt{\frac{8l\left(M l+J \right)}{27G}}+\pi \sqrt{\frac{8l\left(M l-J \right)}{27G}} . 
\end{equation}
It is easy to verify that the first law of black hole thermodynamics,
\begin{equation}
dM=TdS+\Omega \, dJ \, ,
\end{equation}
holds.

The heat capacity at constant angular velocity $C_{\Omega}\equiv T\left(\frac{\partial S}{\partial T} \right)_{\Omega} $ is
\begin{equation}
C_{\Omega}=\frac{16 \pi^2 l^2 T}{27 G\sqrt{1-l^2 \Omega^2}}.
\end{equation}
We note that $C_{\Omega}$ is positive, and therefore the rotating hairy black hole is thermodynamically stable.

\section{Scalar soliton and the black-hole microscopic entropy}\label{sec4}

In this section we will show that, by considering a scalar soliton as the ground state for the case with nonvanishing angular momentum, it is possible to successfully reproduce the hairy black hole entropy by using the  counting of microscopical states approach.  Since we are using the same potential as that in Ref.  \cite{CMT}, we consequently choose the soliton found in this reference as the proper ground state of the rotating hairy black hole. Before explain how this mechanism works, let us briefly review the properties of this scalar soliton. Its metric is given by
\begin{alignat}{2}\notag
ds^2_{\textrm{sol}} &=-\frac{(4l+9r)^{4}}{81l^{2}(8l+9r)^{2}}dt^{2}+\frac{81l^{2}(4l+9r)^{3}%
}{(9r-4l)(8l+9r)^{4}}dr^{2}\\
&+\left( r^{2}-\frac{16}{81}l^{2}\right) d\varphi
^{2}\ ,  \label{soliton asympt}
\end{alignat}
and the scalar field is
\begin{equation}
\phi (r)_{\textrm{sol}}=\mathrm{arctanh}\sqrt{\frac{4l}{8l+9r}}.
\label{scalar field asymtp}
\end{equation}%
The  coordinates range as $r \geq \frac{4 l}{9}$, $-\infty<t<\infty$, $0 <\varphi \leq 2\pi$. The soliton has no curvature singularities and it is  devoid of integration constants. Indeed, its Ricci scalar reads
\begin{equation}
R_{\mu}^{\mu}=-\frac{2(8l+9r)^3(243 r^2 +180 l r+64l^2)}{l^2(4l+9r)^5},
\end{equation}
and since the invariants $R_{\mu \nu}R^{\mu \nu}$ and $R_{\mu \nu \lambda \rho  }R^{ \mu \nu \lambda \rho}$ are regular,  it is possible to conclude that the soliton is smooth and regular everywhere. Additionally, expanding the metric around the origin $r=\frac{4 l}{9}$, followed by the change of the radial coordinate $\rho^2=\frac{8l}{9}\left(r-\frac{4l}{9} \right)$ and with a adequate time rescaling, we get
\begin{equation}
ds^2_{\textrm{sol}} \approx-dt+d\rho^2+\rho^2 d\varphi^2
\end{equation}
 around $\rho=0$, which is the Minkowski spacetime. Thus, we have checked that the soliton has no conical defects. 

The scalar field (\ref{scalar field asymtp}) and metric (\ref{soliton asympt}) of the soliton fit in the same set of relaxed asymptotically AdS conditions of  Refs. \cite{CMT} and \cite{HMTZ-2+1}. This fact allows to compute the soliton charges in the same manner as  for the rotating hairy black hole, yielding
\begin{equation}
M_{\textrm{sol}}=-\frac{2}{27 G} \, ,
\end{equation}
and  a vanishing angular momentum, i.e. the soliton is static.

We are dealing with solutions whose asymptotic behavior belongs to a class having the same asymptotic symmetry group as pure  three-dimensional AdS gravity. In the light of the holographic principle, it is possible to establish a map between the gravity sector and quantum states of a conformal field theory, generated by the Virasoro operators. In a generic case of a two dimensional conformal field theory in a cylinder, the two copies of the zero-mode Virasoro operators are shifted as $\tilde{L}_{0}^{\pm }=L_{0}^{\pm }-\frac{c^{\pm }}{24}$, where $c^{\pm }$ correspond to the central charges. We denote $\tilde{\Delta}^{\pm }$  the eigenvalues of $\tilde{L}_{0}^{\pm }$, meanwhile the lowest ones are given by
\begin{equation}\label{delta}
\tilde{\Delta}_{0}^{\pm }=\Delta_{0}^{\pm }-\frac{c^{\pm }}{24}.
\end{equation}
The entropy coming from the counting of the asymptotic growth of the number of states, so called Cardy formula, can be written as \cite{CMT}
\begin{equation}
S=4\pi \sqrt{-\tilde{\Delta}_{0}^{+}\tilde{\Delta}^{+}}+4\pi \sqrt{-\tilde{%
\Delta}_{0}^{-}\tilde{\Delta}^{-}}\ . \label{Cardy super reloaded}
\end{equation}%
The above expression can be derived directly from Refs. \cite{CarlipPark} and remarkably does not depend explicitly in the central charges. The formula (\ref{Cardy super reloaded}) has been shown to be successful in reproducing the Bekenstein-Hawking entropy for a wide class of non-rotating hairy black holes \cite{CMT,CMT2}. The key point in this mechanism was the assumption that a scalar soliton corresponds to the ground state.

The connection between gravity and the conformal field theory can be
performed associating the mass and the angular momentum with the eigenvalues of the Virasoro operators as follows, 
\begin{equation}\label{rel}
M=\frac{1}{l}(\tilde{\Delta}^{+}+\tilde{\Delta}^{-}), \quad J=\tilde{\Delta}^{+}-\tilde{\Delta}^{-}\, ,
\end{equation}
or conversely $\tilde{\Delta}^{\pm}=\frac{1}{2}\left(M l\pm J \right)$. Let us assume that the ground state is non-degenerated and denote its mass as $M_0$ (naturally, it has no angular momentum). Thus, the lowest eigenvalues are $\tilde{\Delta}^{\pm}_0=l M_0/2$ and we can rewrite (\ref{Cardy super reloaded}) as follows
\begin{equation}
S=2\pi \sqrt{-lM_0\left(M l+J \right)}+2\pi \sqrt{-lM_0\left(M l- J \right)}.
 \label{Cardy super hiper reloaded}
\end{equation}

The energy spectrum of the class of rotating hairy black hole has a continuum sector starting from the zero mass and zero angular momentum solution which coincides with the corresponding in pure gravity (the massless BTZ black hole). Below, separated by a gap, there exist a non-degenerate solution which is the scalar soliton described by Eqs. (\ref{soliton asympt}) and (\ref{scalar field asymtp}). Taking this into account we can evaluate (\ref{Cardy super hiper reloaded}) considering that the ground state is the scalar soliton where $M_0=M_{\textrm{sol}}=-\frac{2}{27G}$, yielding
\begin{equation}\label{Cardy super hiper rotating reloaded}
S=\pi \sqrt{\frac{8l\left(M l+J \right)}{27G}}+\pi \sqrt{\frac{8l\left(M l-J \right)}{27G}} .
\end{equation}
This exactly matches with the Bekenstein-Hawking entropy given in Eq. (\ref{BHentro}).

Following the appendix C in Ref. \cite{CMT}, one can easily derive the entropy in the grand canonical ensamble
\begin{equation} \label{ecs}
S=-\frac{8 \pi^2 l^2 M_{\textrm{sol}} T}{\sqrt{1-l^2 \Omega^2}}= \frac{16 \pi^2 l^2 T}{27 G\sqrt{1-l^2 \Omega^2}} ,
\end{equation} 
which matches with (\ref{BHentro}) once $T$ is expressed in terms of $M$ and $J$ by using Eq. (\ref{temp}).

Some comments are in order. We note that the solutions of the action (\ref{Action}) with the scalar field turned on, cannot smoothly deform into vacuum solutions keeping fixed the mass and the angular momentum. This implies that both hairy and vacuum solutions belong to different and disconnected sectors which have different ground states. The counterpart of the soliton, as the ground state in the hairy sector, is the AdS spacetime in the vacuum sector.  In this case, $\tilde{\Delta_{0}^{\pm }}=l M_{\textrm{AdS}}/2=-l/(16G)$ giving the well-known result for the vacuum sector \cite{Strominger}
\begin{equation}
S_{\textrm{BTZ}}=\pi \sqrt{\frac{l\left(M l+J \right)}{2G}}+\pi \sqrt{\frac{l\left(M l-J \right)}{2G}} .
\end{equation}

\section{Concluding remarks}\label{sec5}

We have determined the entropy of a new rotating hairy black hole using the Cardy formula. The crucial point relies on the identification of a scalar soliton as the ground state of theory. We have assume that there exist an \emph{unique} soliton for each self-interaction potential, and that its mass is associated with the lowest eigenvalues of the zero-mode Virasoro operator $\tilde{L}_0^\pm$. 

Improper gauge transformations in the plane $t-\varphi$ allow to obtain rotating solutions from static ones. Using this idea we have construct a spinning black hole where the improper change of coordinates corresponds to a  Lorentz transformation. The rotating black hole studied here has a single horizon and is unable to become extreme keeping the scalar turned on. Nevertheless, one could consider different improper gauge transformations in such a way that they could produce rotating hairy black holes with a additional (inner) horizon, like the rotating BTZ black hole.  

Static scalar hairy black holes in three dimensions have been found considering self-interaction potentials with one \cite{HMTZ-2+1,CMT} and two coupling constants \cite{CMT2}. It is interesting to explore the existence of rotating black holes following the procedure explained above taking into account these static black holes. Moreover, each such a potential has associated his own scalar soliton. Thus, one could test Cardy formula in the case of more generic rotating hairy black holes. However, we think that the example presented here, which is the simplest one,  reveals the main aspects of this problem.

The derivation of the entropy by the Cardy formula in this paper takes advantage of the asymptotic symmetries and also the holographic principle. On other hand, although we know what states Cardy formula counts in the conformal field theory, this picture does not shed enough light about the specific degrees of freedom in the gravity side. Thus, the problem is opened to further investigations and alternative approaches.

\acknowledgments The authors thank Ricardo Troncoso for the valuable and kind help along this project. F.C. wishes to thank the kind hospitality at Universit\'e de Mons, especially to Mauricio Valenzuela and Per Sundell. This work has
been partially funded by the following Fondecyt grants: 1121031, 1095098, 1100755, and by the Conicyt grants
 79112034 and ACT-91.  The
Centro de Estudios Cient\'{\i}ficos (CECs) is funded by the Chilean Government
through the Centers of Excellence Base Financing Program of Conicyt.

\end{document}